\pdfoutput=1
\documentclass[journal]{IEEEtran}

\usepackage[T1]{fontenc}
\usepackage[utf8]{inputenc}
\usepackage{cite}
\usepackage{graphicx}
\usepackage{amsmath,amssymb}
\usepackage{booktabs}
\usepackage{threeparttable}
\usepackage{array}
\usepackage{placeins}
\usepackage[hidelinks]{hyperref}
\usepackage{orcidlink}
\usepackage{xurl}

\newcommand{\pcm}{\,\mathrm{p/cm^2}}
\newcommand{\opfl}{\Phi_{\mathrm{op}}}
\newcommand{\totfl}{\Phi_{\mathrm{total}}}

\makeatletter
\newenvironment{manuscriptabstract}{%
  \normalfont\@IEEEabskeysecsize\bfseries
  \textit{\abstractname:}\nobreakspace\relax\@IEEEgobbleleadPARNLSP
}{%
  \relax\vspace{1.34ex}\par\normalfont\normalsize
}
\newenvironment{manuscriptkeywords}{%
  \normalfont\@IEEEabskeysecsize\bfseries
  \textit{\IEEEkeywordsname:}\nobreakspace\relax\@IEEEgobbleleadPARNLSP
}{%
  \relax\vspace{0.67ex}\par\normalfont\normalsize
}
\makeatother

\hypersetup{
  pdftitle={Proton Irradiation Characterization of an Open-Source ML Accelerator on a Zynq UltraScale+ MPSoC},
  pdfauthor={Saad Memon, Rafal Graczyk, Jan Swakon, Leszek Grzanka, Sebastian Kusyk, and Mike Papadakis},
  pdfsubject={Proton irradiation of a Linux-managed COTS FPGA-SoC neural-network platform},
  pdfkeywords={COTS, FPGA-SoC, Linux, neural-network accelerator, proton irradiation, silent data corruption, single-event functional interrupt}
}

\begin{document}

\title{Proton Irradiation Characterization of an Open-Source ML Accelerator on a
Zynq UltraScale+ MPSoC}

\author{Saad~Memon~\orcidlink{0000-0001-5672-7573},
        Rafal~Graczyk~\orcidlink{0000-0003-4570-3431},
        Jan~Swako\'n~\orcidlink{0000-0001-9262-7326},
        Leszek~Grzanka~\orcidlink{0000-0002-3599-854X},
        Sebastian~Kusyk~\orcidlink{0000-0001-8915-5313},
        and~Mike~Papadakis~\orcidlink{0000-0003-1852-2547},~\IEEEmembership{Member,~IEEE}%
\thanks{This work has been submitted to the IEEE for possible publication. Copyright may be transferred without notice, after which this version may no longer be accessible. This work was supported in part by the European Union's Horizon 2020
research and innovation programme under Grant Agreement No.~101008126
(RADNEXT), the Horizon Europe research and innovation programme under Grant
Agreement No.~101057511 (EURO-LABS), and the Luxembourg National Research
Fund (FNR) under Grant CS20/IS/14689454 (HERA). For the purpose of open
access, the authors have applied a Creative Commons Attribution 4.0 (CC BY 4.0)
license to any Accepted Manuscript version arising from this submission.}
\thanks{Saad Memon (Corresponding author) and Mike Papadakis are with the Interdisciplinary Centre for
Security, Reliability and Trust (SnT), University of Luxembourg, L-1855
Luxembourg, Luxembourg (e-mail: mail.saadmemon@gmail.com;
michail.papadakis@uni.lu).}
\thanks{Rafal Graczyk was with the Interdisciplinary Centre for Security,
Reliability and Trust (SnT), University of Luxembourg, L-1855 Luxembourg,
Luxembourg.}
\thanks{Jan Swako\'n, Leszek Grzanka, and Sebastian Kusyk are with the Henryk
Niewodnicza\'nski Institute of Nuclear Physics, Polish Academy of Sciences
(IFJ PAN), 31-342 Krak\'ow, Poland (e-mail: jan.swakon@ifj.edu.pl;
leszek.grzanka@ifj.edu.pl; sebastian.kusyk@ifj.edu.pl).}}

\markboth{Preprint. Submitted to IEEE Transactions on Nuclear Science}%
{Memon \MakeLowercase{\textit{et al.}}: Proton Irradiation Characterization of an
Open-Source ML Accelerator on a Zynq UltraScale+ MPSoC}
\maketitle

\begin{manuscriptabstract}
As spaceborne computing systems increasingly rely on neural network (NN)
accelerators, the opacity of commercial, black-box architectures severely
restricts the development of verifiable radiation mitigation strategies.
Open-source, register-transfer level (RTL)-accessible accelerators resolve this
limitation by enabling user-defined instrumentation, yet few have empirical radiation-response baselines. This work establishes a foundational
system-level proton-irradiation baseline for an unmitigated open-source Tensil
NN accelerator deployed on a Zynq UltraScale+ SoC executing ResNet-20
inference. Under 20--58-MeV proton irradiation, we delivered $4.29\times10^{10}$ p/cm$^2$ within monitored operational
windows. Seven workload
interruptions required two restarts of the notebook process, four reboots or board resets, and
one power-cycle sequence. Two output-corruption events returned incorrect
CIFAR-10 classes without loss of service. In the longer event, the accelerator
returned a class absent from the ten-image CIFAR-10 pool for 39 consecutive
inputs at normal cadence. The process remained alive, while the
kernel log, limited memory test, and sampled power showed no anomaly.
Observation of the stuck-class sequence ended with scheduled bitstream
reconfiguration. All
nine onsets occurred under the nominal
4-cm beam, which exposed the SoC, LPDDR4, and additional board circuitry; none
occurred under the 2-cm SoC-centered field. This pattern shows a field association but does not establish LPDDR4 as the cause because field size was confounded with
run order and dose. Linux-managed accelerators require
end-to-end content checks and recovery that reaches the state in which corruption can persist. This baseline documents availability loss
and silent output corruption, supporting future software hardening of
COTS FPGA-SoCs for neural-network inference in space systems.
\end{manuscriptabstract}

\begin{manuscriptkeywords}
Commercial off-the-shelf (COTS), FPGA-SoC, Linux, neural-network accelerator,
proton irradiation, silent data corruption, single-event functional interrupt.
\end{manuscriptkeywords}

\section{Introduction}
\label{sec:introduction}

\IEEEPARstart{A}{n} inference service on a commercial off-the-shelf (COTS)
field-programmable gate array (FPGA) system-on-chip (SoC) is not the
accelerator alone. In a typical deployment, the processing system (PS) runs
Linux and the application, the programmable logic (PL) implements the
accelerator, and external dynamic random-access memory (DRAM) holds the model
and data in flight. Every inference crosses all three. Radiation can
disturb any of them, and at the workload level the disturbance surfaces in one
of two ways. Either the service stops (a process terminates, Linux hangs, a
reset is needed), or it keeps running and delivers a wrong answer.
Operationally, these failure modes are distinct. A liveness watchdog can detect
a stopped workload but cannot verify a completed inference. A content checker
can detect a wrong result while Linux remains responsive. Grouping both under
one generic ``failure'' hides which monitor would have
caught it, which recovery it needs, and which exposure belongs in its rate
denominator.

Campaigns on Linux-managed Zynq platforms have recorded crashes and corrupted
inferences in the same beam session
\cite{agiakatsikas2024,sabogal2019recon}, but their reporting conventions
differed. Agiakatsikas \emph{et al.} classified runs by outcome
\cite{agiakatsikas2024}; Sabogal \emph{et al.} grouped consecutive erroneous
outputs as one error \cite{sabogal2019recon}; and Stirk \emph{et al.}
power-cycled after each Linux-benchmark failure \cite{stirk2023}. None reported
how long a corrupted state persisted or what the standard monitors recorded
during the event. Device-level cross sections for configuration
memory, block RAM, and processor resources \cite{hiemstra2017,stirk2023} do
not resolve this system-level question because utilization, activation, and
masking stand between an upset and a wrong answer. The open question is not
whether a Linux-managed accelerator can silently produce wrong outputs; prior
work shows that it can. Instead, we ask what happens when corrupted state
persists through a block of inputs: how long it lasts, how it ends, and what
the available operational monitors report during the event.

\begin{figure}[!t]
\centering
\includegraphics[width=\columnwidth,trim=6pt 2pt 6pt 2pt,clip]{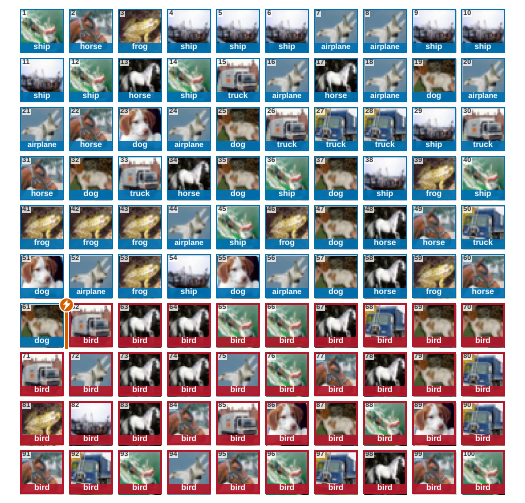}
\caption{Run~5, block~10 (Section~\ref{sec:signatures}), in execution order: 61 correct predictions, then 39 consecutive predictions of \texttt{bird}, a class absent from the ten-image pool, while every availability indicator stayed nominal. Composite of CIFAR-10 test-set images 10--19 \cite{cifar}.}
\label{fig:cifar-collapse-overview}
\end{figure}

To investigate this question, we conducted a proton-irradiation campaign on an
Avnet Ultra96-V2, a Zynq UltraScale+ XCZU3EG with adjacent LPDDR4, running PYNQ
Linux and the open-source Tensil
Tensor Compute Unit (TCU) in the PL \cite{tensil}. The TCU executed a
ResNet-20 classifier on a fixed pool of ten CIFAR-10 images
\cite{resnet,cifar} in blocks of 100 inferences, and the PL was reloaded at each
block boundary. We retained every predicted class in execution order and
monitored workload progress, the kernel log, network reachability, sampled
power rails, and a userspace memory test. In one block at 58\,MeV, the first 61
predictions were correct and the next 39 were all \texttt{bird}, a class absent
from the pool (Fig.~\ref{fig:cifar-collapse-overview}). These outputs arrived at
the usual cadence while every availability indicator remained nominal, until
the scheduled reload ended the observation. The workload was also interrupted
seven times, with recovery ranging from a process restart to a power cycle.
We used two beam fields. One nominally covered the SoC, whereas the wider field
also covered the LPDDR4 package and surrounding board area. This geometry
enabled an exploratory comparison between SoC-focused and wider-system
exposure.

We pose two questions. \emph{Q1:} When corrupted state persists through a
block, how long does it last, how does it end, and what do the Linux liveness,
timing, memory-test, and sampled-power indicators report? \emph{Q2:} Does exposing board circuitry beyond the SoC footprint
change the rate of Linux-level workload interruptions at matched beam
energies?

From this monitored PS--PL inference stack, we report a silent constant-class
corruption and its complete monitor record, seven Linux-level interruptions and
their recovery tiers, and condition-specific cross sections for both endpoints.
We also report a matched-energy field comparison, but field is confounded with
run order and dose. We neither identify a physical mechanism nor
evaluate a mitigation. Instead, we apply and recommend three reporting
practices. First, count events by onset so consecutive wrong outputs from one
corrupted state form one event. Second, match each denominator to the monitor
that observes its endpoint and state that monitor's blind spots. Third, record
the recovery that restored operation without treating it as evidence of the
fault's origin.

\section{Related Work and Research Gap}
\label{sec:related}

\subsection{Device, Platform, and Accelerator Campaigns}

Device-level irradiation of Zynq UltraScale+ and related SRAM-based FPGA
families has quantified susceptibility in configuration memory, block RAM,
flip-flops, and digital-signal-processing (DSP) blocks \cite{hiemstra2017}.
Platform-level campaigns have added processor-side cache and
translation-lookaside-buffer (TLB) cross sections
\cite{stirk2023,agiakatsikas2024}.

Stirk \emph{et al.} compared neutron-test methods for an
Ultra96 and reported process/hang, kernel, and application-processing-unit
reset failures under Linux \cite{stirk2023}. Agiakatsikas \emph{et al.}
irradiated a Linux-managed ZCU102 running a Vitis deep-learning processing
unit (DPU) and observed both crashes and numerical corruption
\cite{agiakatsikas2024}. Sabogal \emph{et al.} irradiated a Linux-managed
convolutional-neural-network (CNN) accelerator on Zynq-7020 and ZU3EG boards
with wide-spectrum neutrons \cite{sabogal2019recon}. They recorded erroneous
and hung executions and grouped consecutive erroneous outputs as one error,
but did not report an episode's duration or its monitor record. Running under
an operating system also changes which
failures appear: Santini \emph{et al.} found that Linux raised the
functional-interrupt rate of an embedded SoC under neutrons without a
comparable change in its silent-corruption rate \cite{santini2016os}.

Radiation and fault-injection studies of FPGA accelerators have examined
quantization, architectural parallelism, redundancy, and scrubbing
\cite{rech2024survey,libano2020quantization,maillard2023,souvatzoglou2024qnn,benevenuti2025finn}.
The reported outcome may be a changed class, a numerical deviation, a hang, or
an accelerator reset. It depends on the checker: a top-1 oracle observes
only class-changing corruption, whereas a score-vector checker detects
numerical changes that preserve the winning class. The protocol matters as
well. Reloading the PL after every input ends any persistent error that another
campaign might observe across many inputs. For example, a Linux-managed Jetson
SoC showed persistent output errors when its main memory was inside the beam
\cite{badia2025vit}. FINN-generated
designs have beam baselines \cite{souvatzoglou2024qnn,benevenuti2025finn};
we found no published beam data for Tensil, and beam data for open
accelerators exercised through a Linux-managed PS--PL path remain scarce.
We treat the complete inference path as the service. This path extends from PL
configuration through direct-memory-access (DMA) control and the Advanced
eXtensible Interface (AXI) interconnect to external memory. The open
register-transfer-level (RTL) source permits future rebuilding and
instrumentation. Because this campaign added neither instrumentation nor
mitigation, its results provide an unhardened baseline for that work.

\subsection{Attribution Under Whole-System Exposure}

External DRAM creates a recurring attribution problem. A wide beam exposes the
SoC, memory package, traces, clocks, and power-delivery circuitry at the same
time. System-level observations alone cannot identify which component
contained the initiating disturbance. Guertin and Cui noted this difficulty
for a SoC tested with its supporting LPDDR memory \cite{guertincui2017}. Our
proton tests of two Linux-managed LPDDR platforms also could not attribute
memory-test mismatches to a component under whole-system exposure
\cite{memon2025qrs}. Those tests used 20--58\,MeV at the device from a nominal
60-MeV beam. Following system-level single-event-effect (SEE) guidance, we use
field geometry only to test association at matched energies. We do not infer a
physical fault location from a workload symptom alone
\cite{guertin2019,quinn2014}.

\section{Experimental Method}
\label{sec:method}

\subsection{Hardware and Software Stack}
\label{sec:platform}

The device under test (DUT) was an Avnet Ultra96-V2 populated with an
XCZU3EG-1SBVA484I Zynq UltraScale+ multiprocessor SoC (MPSoC). The device uses a 16-nm FinFET
process, speed grade $-1$, and an industrial temperature rating. The board
contained 2\,GB of Micron MT53D512M32D2DS-053 LPDDR4 on the PS DDR
controller \cite{ultra96hw} in a 32-bit configuration without error-correcting
code (ECC), which the single x32 device cannot provide \cite{ug1085}. A 16-GB
microSD card held the boot image and root filesystem. The device lot and date
codes were not recorded.

A Rigol DP821 supplied 8.000\,V, within the board's 7--14\,V input range
\cite{ultra96hw}, with remote sensing. Overvoltage and
overcurrent limits were 8.4\,V and 3.5\,A, respectively. The heat sink was
removed to expose the package, and the SoC surface was oriented approximately
normal to the beam. The DUT operated in room-temperature air.

PYNQ Linux v2.7, based on Ubuntu~20.04 with kernel
\texttt{5.4.0-xilinx-v2020.2}, ran on the PS\@. A laptop outside the irradiation
room controlled the board through USB-gadget Ethernet. It launched notebook
blocks and collected outputs and kernel messages. It also sent Internet Control
Message Protocol (ICMP) echo requests to assess reachability. All Linux evidence reached the laptop through this link, which limits what
service loss alone can establish (Section~\ref{sec:limitations}). The platform used
no experiment-specific watchdog, configuration readback, Soft Error Mitigation
(SEM) controller, or redundant inference checker \cite{pg187}.

The inference workload was a ResNet-20 classifier for CIFAR-10. The design was
adapted from the public Tensil Ultra96-V2 flow \cite{tensil}. The TCU used a
$16\times16$ systolic array at 100\,MHz and the 16-bit fixed-point
\texttt{FP16BP8} format throughout. This format covered the array, local memory,
accumulator memory, and single-instruction, multiple-data (SIMD) datapath. The
TCU contained 20,480 local-memory vectors and 4,096
accumulator-memory vectors. For the compiled model used in this campaign,
Tensil reported 27 layers, 101,840 nine-byte instructions, 568,474 constant
scalars, and 97.2\% constant utilization.

Two 128-bit AXI4 master interfaces moved
weights, inputs, intermediate data, and outputs between the TCU and external
LPDDR4\@. A 128-bit AXI direct-memory-access (DMA) path supplied TCU
instructions. Local and accumulator memories resided in PL block
RAM\@. Each inference exercised PS software, PS--PL interfaces, external
LPDDR4, DMA and interconnect logic, configuration state, and accelerator-local
state.

The bitstream used in the campaign was implemented with Vivado~2021.2 using
the original RTL, target device, and board definition. The routed design met
timing with $+0.523$\,ns worst-case slack. It used
24,382 lookup tables (34.6\%), 18,907 registers (13.4\%), 180 block-RAM tiles
(83.3\%), and 273 DSP48E2 blocks (75.8\%).

Vivado's essential-bit mask for this design marks
$B_{\mathrm{ess}}=13{,}222{,}948$ of 30,876,800 positions (42.8\%) as
essential. We report this count for future comparison but do not use it for
normalization.

\subsection{Inputs, Oracle, and Observability}

The input pool contained zero-based indices 10--19 of the canonical CIFAR-10
test set. These ten images span six classes: truck, dog, horse, and ship each
appear twice, while airplane and frog each appear once. Automobile, bird, cat,
and deer are absent. A pre-irradiation validation classified all ten images correctly, so the
dataset labels served as the online reference.

Each inference selected one image uniformly with replacement. No fixed
pseudorandom seed was used, so the realized order cannot be regenerated from
the sampling rule alone. The log retained the selected input index, reference
class, and predicted class for every retained record.

The checker retained only the predicted class, defined by the largest output
score. It did not retain the complete ten-element score vector, intermediate
tensors, or accelerator state. It can detect a class change but not
a numerical perturbation that leaves the winning class unchanged. We considered
a block verified when its class summary was retained, with or without the
per-inference order. Of the 92 launched blocks, 62 met this criterion. All
availability analyses use a separate operational-window denominator that does
not assume output verification for missing blocks.

\subsection{Irradiation Conditions}
\label{sec:beam}

The AIC-144 cyclotron at IFJ PAN delivered nominal 20-, 40-, and 58-MeV
protons in air \cite{swakon2010}. The beam was normal to the exposed board
surface. Facility dosimetry followed the station procedure, with 3\% reported
uncertainty for fluence and flux. Dose-to-fluence conversion used SRIM2013
stopping powers \cite{srim}. We report nominal incident energy because
component-specific energy was not measured. The three energies are settings
below the facility's nominal 60-MeV proton beam.

\begin{figure}[!t]
\centering
\includegraphics[width=\columnwidth]{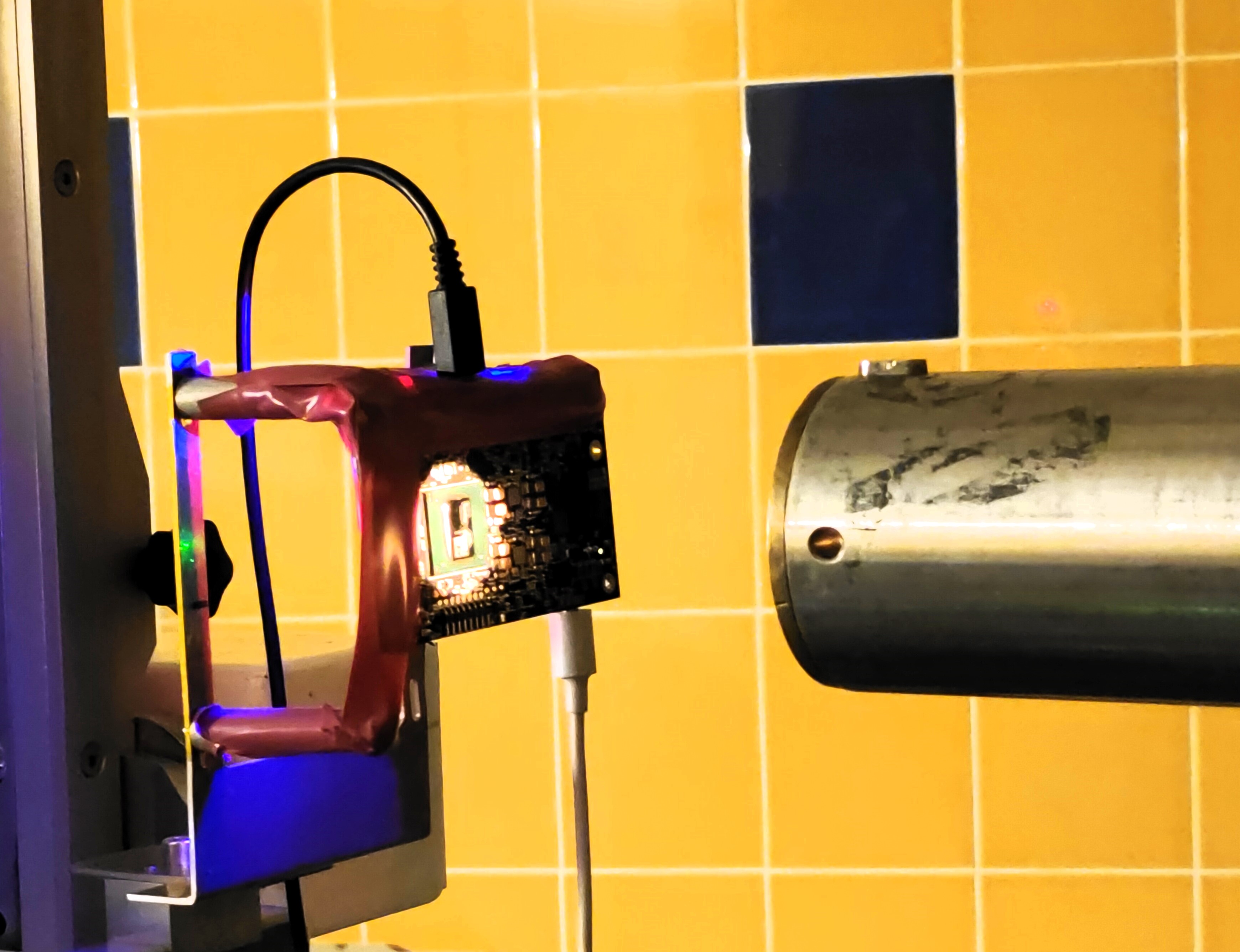}
\caption{Ultra96-V2 DUT at the AIC-144 station with the heat sink removed and the SoC facing the beam collimator (right).}
\label{fig:dut}
\end{figure}

Two collimated fields changed the exposed board area; Table~\ref{tab:beam}
labels them by the operator's nominal designation. The operator record
describes the nominally 2-cm field as primarily covering the SoC\@. The
facility record identifies a 25-mm collimator, which we use as the
authoritative aperture. Because this exceeds the 19-mm SBVA484 package, the
edge of the adjacent LPDDR4 package may also have been exposed. The nominally
4-cm field covered the SoC,
the LPDDR4 package, and additional board area. Both fields were used at 20 and
40\,MeV. Only the wider field was used at 58\,MeV. Fig.~\ref{fig:dut} shows
the mounted DUT\@. Table~\ref{tab:beam} lists the six
monitored irradiation intervals. The monitored intervals accumulated
$6.19\times10^{10}\pcm$.

\begin{table}[!t]
\caption{Proton Beam Conditions for Monitored Runs}
\label{tab:beam}
\centering
\begin{threeparttable}
\footnotesize
\setlength{\tabcolsep}{2.7pt}
\begin{tabular*}{\columnwidth}{@{\extracolsep{\fill}}ccccccc@{}}
\toprule
Run & $t$ & $E$\tnote{a} & Flux & $\totfl$ & $D_w$ & Field\\
& (s) & (MeV) & (p/cm$^2$/s) & (p/cm$^2$) & (Gy) & (cm)\\
\midrule
1  & 1126.5 & 20 & $1.14\times10^{7}$ & $1.29\times10^{10}$ & 54.7 & 2\\
2  & 1076.4 & 40 & $1.20\times10^{7}$ & $1.29\times10^{10}$ & 31.2 & 2\\
3  & 1016.4 & 20 & $1.01\times10^{7}$ & $1.02\times10^{10}$ & 43.4 & 4\\
4a & 781.4  & 40 & $1.20\times10^{7}$ & $9.37\times10^{9}$  & 22.6 & 4\\
4b & 388.2  & 40 & $1.28\times10^{7}$ & $4.98\times10^{9}$  & 12.0 & 4\\
5  & 1191.8 & 58 & $9.61\times10^{6}$ & $1.15\times10^{10}$ & 20.6 & 4\\
\midrule
Total & 5580.7 & & & $6.19\times10^{10}$ & 184.5 & \\
\bottomrule
\end{tabular*}
\begin{tablenotes}\scriptsize
\item[a] $t$ is irradiation time, $E$ is nominal incident energy, and $D_w$ is absorbed dose to water. Facility dose, fluence, and flux uncertainty is 3\%. Run~4 contains intervals 4a and 4b; unmonitored exposures are excluded.
\end{tablenotes}
\end{threeparttable}
\end{table}

\subsection{Block Protocol and Recovery}
\label{sec:procedure}

One block corresponded to one notebook execution. At block start, the
notebook issued two PL configuration downloads, ran one
\texttt{memtester 10M 1} pass (a stuck-address test and 17 pattern tests over a
10-MB region, 0.5\% of the memory), and sampled the board's power rails
\cite{memtester}. The notebook then executed and logged 100 inferences, sampled
the power rails again, recorded a class summary, and performed a final PL
configuration download.

The 100-inference sequence required about 2.34\,s. Individual output records
were typically 23--24\,ms apart. The userspace memory test typically required
15--20\,s. Successive block starts were separated by a median of 40\,s, with a
23--127\,s range.

The final reconfiguration rewrote PL configuration, reinitialized block RAM,
reset the TCU and DMA, and reloaded the model, which rewrote the model
constants in their LPDDR4 buffer. It did not reset Linux, the PS, on-chip
memory (OCM), the remaining LPDDR4 contents, or the Linux page cache.

Operators continuously followed block summaries, the control terminal, kernel
messages, and network reachability. A continuous supply-current trace is
available only for run~1; the other runs contain two power-rail samples per
block. After a sustained loss of progress, recovery began with the least
disruptive available action. Operators escalated through process restart,
Linux reboot, board reset, and power cycle. When several actions changed
together or observation ended, we report a lower bound or the full recovery
sequence rather than assign causality to one action.

\subsection{Event Definitions and Statistical Analysis}
\label{sec:defs}

Individual particle interactions were not observed directly. We count an
\emph{operational event onset} as the first abnormal observation after a
recorded normal state or recovery action. Consecutive abnormal outputs, log
records, or failed relaunches without an intervening normal state form one
cluster. A recorded process restart, reboot, reset, or power cycle ends the
cluster, so a later failure begins a new onset. One persistent corruption is
therefore counted once, regardless of how many predictions it affects.
Clustering is performed separately for each endpoint. Thus, an output event and
the interruption that follows it are counted under their respective endpoints.
This rule affects run~4, whose record is ambiguous. The primary count treats a
recorded reboot as a cluster boundary although no block completed before the
next failure; Section~\ref{sec:partition} reports alternative readings.

We retain the label numbering of the campaign's event taxonomy and define
only the categories used here. F3 denotes multiple incorrect predictions
followed by block completion, and F4 denotes output corruption still present
when observation ended. F6 denotes an isolated accelerator or DMA timeout;
F7, an isolated PS--PL interface failure; F8, failure of the prescribed user
process; and F9, a Linux- or system-level failure. The labels describe observed behavior, not
fault locations. Recovery levels identify the least disruptive successful
action: R5 is a process restart, R6 or R7 a reboot or board reset, and R8 a
power cycle.

Following the workload-conditioned definition of our prior study, we classify
every F8 or F9 onset that prevented completion of the prescribed notebook
workload as a Linux-level single-event functional interrupt (Linux-SEFI)
\cite{memon2025}. Its tiers correspond to the process, kernel, and reset classes
of Stirk \emph{et al.} \cite{stirk2023} and the symptom classes of Esquer
\emph{et al.} \cite{esquer2024sefi}. The term describes an observed symptom: a
workload that stopped and required at least a process restart. It does not
identify a location or mechanism. The definition includes both process failure
while Linux remains reachable and any stall observed as a loss of progress
(Section~\ref{sec:signatures}). The count can therefore include interruptions
of any origin, so each Linux-SEFI cross section is an upper bound on the
radiation-induced rate. Unlike the JEDEC SEFI definition \cite{jesd89a}, our
definition includes the run~4 episode that ended in a power cycle.
Table~\ref{tab:counts} lists this episode separately so it can be excluded.

We report two primary endpoints separately: every F3/F4 inference-output
onset and every F8/F9 Linux-SEFI onset. Corrected-OCM reports (OCM ECC
correctable-error reports from the Linux driver) are telemetry rather than
workload failures and form a separate endpoint. By itself, such a report
establishes neither radiation origin nor the software object at that address.
Every onset began while the beam was on, so we call them
radiation-associated. This criterion is temporal. The error-free pre-beam
blocks and the 44 completed small-field blocks, including 31 with verified
outputs, weigh against but do not exclude a beam-independent cause.

The Linux-SEFI denominator is operational-window fluence, denoted
$\Phi_{\mathrm{op}}$. For each contiguous monitored window, we integrate the
facility-reported flux from the first block launch until the final block ends
or failure recovery begins. Because the beam remained on between blocks,
$\Phi_{\mathrm{op}}$ includes inter-block intervals; it is not the fluence of
the 2.34-s inference phases alone. Classification evidence was retained for
only 62 of 92 launched blocks. For output events, we therefore allocate each
run's operational fluence in proportion to the fraction of blocks with
retained output verification:
\begin{equation}
\Phi_{\mathrm{ver}}^{(r)}=\Phi_{\mathrm{op}}^{(r)}
\frac{n_{\mathrm{ver}}^{(r)}}{n_{\mathrm{blk}}^{(r)}},
\label{eq:phiver}
\end{equation}
where $n_{\mathrm{ver}}^{(r)}$ and $n_{\mathrm{blk}}^{(r)}$ are verified and
launched block counts. Equation~\eqref{eq:phiver} is an exposure-allocation
approximation: it does not assert that verified and unverified blocks share
the same latent event history.

For endpoint $k$ under condition $g$, the observed system-level event cross
section is
\begin{equation}
\sigma_{k,g}=
\frac{\displaystyle\sum_{r\in g}N_{k,r}}
{\displaystyle\sum_{r\in g}\Phi_{k,r}}
\quad [\mathrm{cm^2/system}],
\label{eq:sigma}
\end{equation}
where $N_{k,r}$ is the number of clustered onsets and $\Phi_{k,r}$ the
endpoint-specific denominator; the subscripts SEFI and FE denote the
Linux-SEFI and the F3/F4 output-event endpoints. We write cm$^2$/system for a
cross section per DUT under the stated field; because the exposed area differs
between fields, the values are condition-specific. For nonzero counts we
report two-sided 95\% Garwood intervals under a Poisson counting model, and
for zero observed events the one-sided 95\% upper limit
\begin{equation}
\sigma_{k,g}^{95\%,\mathrm{UL}}=
\frac{2.996}{\displaystyle\sum_{r\in g}\Phi_{k,r}},
\label{eq:zero_ul}
\end{equation}
where $2.996=-\ln 0.05$. These intervals quantify counting uncertainty only
and assume independent onsets at a constant rate within each window. Clustering
or a rate that changes with accumulated dose would widen the intervals. When a
window ends at the failure that terminated the run, as in run~4, the ratio
$N/\Phi_{\mathrm{op}}$ is biased upward for small counts. The intervals do not
include this bias.

The facility assigns 3\% uncertainty to fluence and flux. Minute-resolution
operator timestamps add reconstruction bounds of $\pm90$\,s for runs~1--2 and
$\pm150$\,s for runs~3--5. These bounds equal 8.6\%, 9.2\%, 19\%, 62\%, and
17\% of the reconstructed $\Phi_{\mathrm{op}}$ for runs~1--5. Because these are
bounds rather than independent random errors, we do not combine them in
quadrature with dosimetry uncertainty. Instead, we repeat the field comparison
after lengthening every wide-field window and shortening every small-field
window by its bound. This adjustment weakens the observed contrast.

For two conditions modeled as independent Poisson counts, conditioning on the
total count under equal rates per fluence yields a binomial allocation. If
conditions 1 and 2 have exposures $\Phi_1$ and $\Phi_2$, the null probability
that an event falls in condition~1 is
\begin{equation}
p_0=\frac{\Phi_1}{\Phi_1+\Phi_2}.
\label{eq:conditional_p}
\end{equation}
For the matched-energy comparison, we condition on the total count within each
energy and use the total wide-field count as the test statistic. When every
event falls in the wider field, the one-sided exact $p$-value is the product of
the per-energy conditional probabilities. The analysis was not preregistered,
so these $p$-values provide exploratory rate evidence rather than confirmatory
hypothesis tests. We apply no multiplicity adjustment.

\section{Results}
\label{sec:results}

\subsection{Exposure and Data Completeness}

\begin{figure*}[t]
\centering
\includegraphics[width=\textwidth]{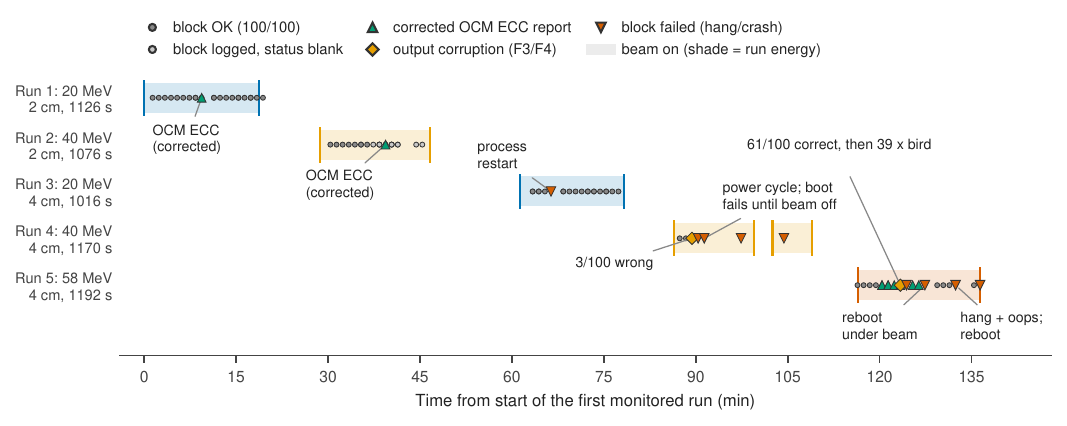}
\caption{Monitored campaign timeline. Open circles mark logged blocks without a recorded class status; a corrected-OCM marker replaces a block's status marker; run~4 comprises intervals 4a and 4b. Annotations reproduce operator notes and event order; they do not establish causality. All seven Linux-SEFI and both output-event onsets began under the wider field.}
\label{fig:timeline}
\end{figure*}

\begin{table}[!t]
\caption{Run Summary}
\label{tab:runsum}
\centering
\begin{threeparttable}
\footnotesize
\setlength{\tabcolsep}{1.0pt}
\begin{tabular*}{\columnwidth}{@{\extracolsep{\fill}}cccccccccc@{}}
\toprule
Run & $E$ & Field & $\opfl$ & \multicolumn{3}{c}{Blocks} & Err. & Event & Op.\\
& (MeV) & (cm) & ($10^{10}$\,p/cm$^2$) & launch. & compl. & verif. & outputs & onsets & time (\%)\\
\midrule
1 & 20 & 2 & 1.19 & 23 & 23 & 23 & 0 & 0 & 93\\
2 & 40 & 2 & 1.17 & 21 & 21 & 8  & 0 & 0 & 91\\
3 & 20 & 4 & 0.78 & 20 & 19 & 19 & 0 & 1 & 76\\
4 & 40 & 4 & 0.29 & 7  & 3  & 3  & 3 & 3 & 21\\
5 & 58 & 4 & 0.86 & 21 & 17 & 9  & 39 & 5 & 75\\
\midrule
$\Sigma$ & & & 4.29 & 92 & 83 & 62 & 42 & 9 & 70\\
\bottomrule
\end{tabular*}
\begin{tablenotes}\scriptsize
\item Blocks are launched, completed, and verified (retained class records); Err.\ outputs are incorrect predictions (affected outputs, not event counts). The nine primary onsets are seven Linux-SEFIs and two output events. Output-event cross sections use $\Phi_{\mathrm{ver}}$ from the 62 blocks with retained class evidence. Op.\ time is $t_{\mathrm{op}}/t$, where $t_{\mathrm{op}}=\opfl/\mathrm{flux}$ is the operational-window duration and $t$ is the run duration in Table~\ref{tab:beam}.
\end{tablenotes}
\end{threeparttable}
\end{table}

The monitored runs delivered $6.19\times10^{10}\pcm$. Reconstructed operational
windows account for $4.29\times10^{10}\pcm$; the remainder was delivered before
the first monitored block, after the last, or during recovery.
Fig.~\ref{fig:timeline} places each window, recovery interval, and onset in
campaign order. Table~\ref{tab:runsum} gives the per-run accounting.
Ninety-two blocks were launched and 83 completed, yielding 8300 inferences.
Class records were retained for 6200 inferences in 62 blocks. Record retention
varied by energy: 42 of 43 launched blocks at 20\,MeV retained their records,
compared with 11 of 28 at 40\,MeV and 9 of 21 at 58\,MeV. Thus, output
verification was thinnest at the energies where output events occurred. In
run~5, the eight completed blocks without class records were also the eight
blocks with corrected-OCM reports (Fig.~\ref{fig:timeline}). We do not know why
these records were lost. If the loss correlated with abnormal blocks,
\eqref{eq:phiver} allocates too much exposure to those runs and underestimates
the output-event cross sections. The zero-event result for the 40-MeV small
field is based on only eight verified blocks. Nine launched blocks did not
complete, two more than the seven Linux-SEFI onsets, because failed relaunches
within one cluster are not separate onsets (Section~\ref{sec:defs}).

\subsection{Silent Output Corruption}
\label{sec:signatures}

\begin{figure*}[t]
\centering
\includegraphics[width=\textwidth]{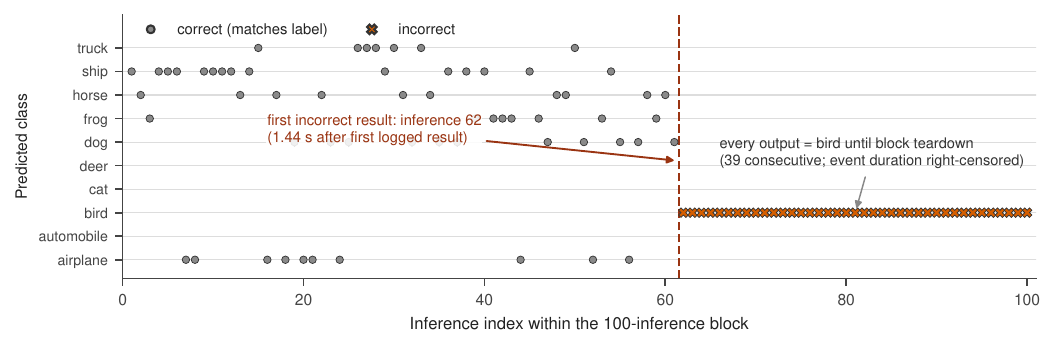}
\caption{Run~5, block~10: 61 correct outputs followed by 39 constant \texttt{bird} outputs. Observation ended at the scheduled block-end reconfiguration.}
\label{fig:collapse}
\end{figure*}

\begin{table}[!t]
\caption{What the Monitors Recorded During the Constant-Class Block}
\label{tab:monitors}
\centering
\begin{threeparttable}
\footnotesize
\setlength{\tabcolsep}{2.5pt}
\begin{tabular*}{\columnwidth}{@{\extracolsep{\fill}}>{\raggedright\arraybackslash}p{0.24\columnwidth}>{\raggedright\arraybackslash}p{0.30\columnwidth}>{\raggedright\arraybackslash}p{0.40\columnwidth}@{}}
\toprule
Indicator & Check and timing & Record for run~5, block~10\\
\midrule
Notebook process & alive, no exception; continuous & alive, no exception\\
Output cadence & interval between records; every inference & 23--24\,ms throughout (each instruction stream completed, no DMA error reported)\\
Kernel log & kernel-reported errors; continuous over the link & no related message (the userspace-driven accelerator path emits none)\\
Memory test (10\,MB) & pattern mismatches; once, before inference & passed\\
Power rails & sampled voltages and currents; twice per block & within campaign range\\
Corrected-OCM telemetry & ECC-corrected OCM reads; logged on read & reports in adjacent blocks; none attributable to this block in the record\\
Top-1 oracle\tnote{a} & predicted class versus reference; every inference & outputs 62--100 wrong (39 consecutive, one absent class)\\
\bottomrule
\end{tabular*}
\begin{tablenotes}\scriptsize
\item All rows except the last are availability or telemetry indicators and do not use a content oracle; none examines what the accelerator computed. \item[a] Added for the experiment; retained for 62 of 92 blocks.
\end{tablenotes}
\end{threeparttable}
\end{table}

In run~5, block~10, the first 61 predictions matched their references and
predictions 62--100 all read \texttt{bird}
(Figs.~\ref{fig:cifar-collapse-overview} and \ref{fig:collapse}). No bird image
exists in the pool, so a stale but valid input cannot explain the sequence.
The constant class (index 2) differs from the last correct output (dog, index
5), excluding a frozen output buffer. It also differs from the first-maximum
argmax of an all-equal score vector (index 0), excluding a zeroed output vector.
A corrupted input or output path remains possible. The same class persisted
across 39 changing inputs, which rules out a single mislabeled input and
establishes a sustained class-output error in the observed sequence. Because
outputs 1--61 were correct and the wrong class did not depend on the input, the
corrupted state became effective between inferences 61 and 62. This transition
occurred within one 23--24\,ms interval of the inference phase, after model load
and PL configuration at block start, and in state not refreshed between
inferences.

Every availability indicator stayed nominal (Table~\ref{tab:monitors}). The
notebook process remained alive and raised no exception. Outputs kept their
23--24\,ms cadence, showing that each instruction stream completed without a
reported DMA error. The kernel log was silent, as expected for corruption
inside a userspace-driven accelerator. The block's memory test had passed, and
its power-rail samples were in range. A watchdog using only these signals would
not have detected the 39 incorrect outputs. The record does not show whether a
corrected-OCM report was logged inside this block; the same address was
reported in the blocks around it (Section~\ref{sec:linuxsefi}).

The error was still present at the block boundary. It did not clear during the
39 affected predictions, about 0.9\,s at nominal cadence. Nothing in the
protocol acted until the scheduled PL reload, which rewrote the configuration,
reinitialized the PL memories, and reloaded the model. The observation
therefore ended with the reload. The protocol, rather
than the fault, right-censored the episode at 39 predictions.

The next launched block did not complete (Fig.~\ref{fig:timeline}), so the
record does not show whether the reload ended the underlying abnormal state.
If the subsequent failure continued the same condition, the output event and
one run~5 Linux-SEFI would be one physical episode counted under two endpoints.
The run~4 output event was also followed by a failed launch: the F3 block at
89.4\,min, then a failure at 90.4\,min (Fig.~\ref{fig:timeline}). The record
cannot determine whether one condition produced both symptoms.
Section~\ref{sec:crosssections} therefore includes a merged reading. We report
one onset and 39 affected outputs, separating the event frequency from its
persistence. Treating all 39 outputs as independent events would overcount the
episode by a factor of 39.

The second output event, in run~4 at nominal 40\,MeV, survives only as a block
summary: three incorrect predictions in a block that completed (F3). Its
output order was not retained, so the record supports one to three distinct
onsets; the primary analysis uses one, and Table~\ref{tab:counts} carries the
uncertainty. No isolated accelerator or DMA timeout (F6) or PS--PL interface
failure (F7) was identified. The instrumentation could not distinguish either
condition because the driver waits for the accelerator without a timeout. A
stalled accelerator or DMA would therefore appear as a loss of workload
progress and be classified as a Linux-SEFI.

\subsection{Availability Loss: Linux-SEFIs, Recovery, and Telemetry}
\label{sec:linuxsefi}

\begin{table}[!t]
\caption{The Nine Primary Onsets in Campaign Order}
\label{tab:onsets}
\centering
\begin{threeparttable}
\scriptsize
\setlength{\tabcolsep}{1.5pt}
\begin{tabular*}{\columnwidth}{@{\extracolsep{\fill}}ccl>{\raggedright\arraybackslash}p{0.42\columnwidth}@{}}
\toprule
Run & Block time (min) & Endpoint & Evidence and recovery\\
\midrule
3 & 66.4 & F8 Linux-SEFI & process termination, no retained signal or kernel message; process restart (R5)\\
4 & 89.4 & F3 output & three wrong outputs, order not retained; block completed; next launch failed\\
4 & 90.4, 91.4 & F9 Linux-SEFI & two failed launches; loss of progress and service; reboot (R6/R7)\\
4 & 97.4, 104.4\tnote{a} & F9 Linux-SEFI & failed relaunch after the reboot, then boot failures under beam; beam off and power cycle (R8)\\
5 & 123.4 & F4 output & 61 correct, then 39 consecutive \texttt{bird}; censored at the reload; next launch failed\\
5 & 124.4 to 136.4\tnote{b} & F8 + 3 F9 Linux-SEFIs & one segmentation fault (R5); three reboot-tier interruptions, one with a photographed oops, one recovery-censored (R6 lower bound)\\
\bottomrule
\end{tabular*}
\begin{tablenotes}\scriptsize
\item Times are block launches in Fig.~\ref{fig:timeline}. \item[a] The 104.4-min launch lies in interval 4b, inside the recovery sequence. \item[b] Failed launches at 124.4, 127.4, 132.4, and 136.4\,min; the timeline places the kernel oops at the final launch; the archive does not
map the other launches to onsets.
\end{tablenotes}
\end{threeparttable}
\end{table}

\begin{table}[!t]
\caption{Raw Observed Outcomes by Nominal Energy}
\label{tab:counts}
\centering
\begin{threeparttable}
\footnotesize
\setlength{\tabcolsep}{2.1pt}
\begin{tabular*}{\columnwidth}{@{\extracolsep{\fill}}>{\raggedright\arraybackslash}p{0.62\columnwidth}ccc@{}}
\toprule
& 20\,MeV & 40\,MeV & 58\,MeV\\
\midrule
Primary operational onsets\tnote{a} & 1 & 3 & 5\\
\quad multi-output deviation (F3)\tnote{b} & 0 & 1 & 0\\
\quad sustained constant-class output (F4) & 0 & 0 & 1\\
\quad process-restart Linux-SEFI (F8, R5) & 1 & 0 & 1\\
\quad reboot/reset-tier Linux-SEFI (F9)\tnote{c} & 0 & 1 & 3\\
\quad power-cycle-sequence Linux-SEFI (F9, R8) & 0 & 1 & 0\\
Corrected-OCM telemetry episodes & 1 & 1 & 1\\
Observed incorrect inference outputs & 0 & 3 & 39\\
Launched blocks not completed & 1 & 4 & 4\\
\bottomrule
\end{tabular*}
\begin{tablenotes}\scriptsize
\item[a] F3/F4 output events plus F8/F9 Linux-SEFIs. \item[b] Output order was not retained; one onset is the minimum supported count. \item[c] Includes one recovery-censored run~5 onset. Energy columns mix field conditions and are not an energy-response experiment.
\end{tablenotes}
\end{threeparttable}
\end{table}

\begin{table}[!t]
\caption{Observed Linux-SEFI Evidence and Recovery}
\label{tab:linuxsefi}
\centering
\begin{threeparttable}
\footnotesize
\setlength{\tabcolsep}{2.5pt}
\begin{tabular*}{\columnwidth}{@{\extracolsep{\fill}}>{\raggedright\arraybackslash}p{0.20\columnwidth}
c >{\raggedright\arraybackslash}p{0.66\columnwidth}@{}}
\toprule
Recovery category & $N$ & Runs and available evidence\\
\midrule
Process restart (R5) & 2 &
Run~3: Jupyter/IPython process termination. Run~5: segmentation fault with a
core dump of the notebook server process. \\
\addlinespace[1pt]
Reboot/reset (R6/R7) & 4 &
One event in run~4 and three in run~5. Evidence included workload-progress
loss, notebook or secure-shell service loss, and a photographed null-pointer
kernel oops with a call trace in the \texttt{clone}/fork path while
\texttt{ipython} was active. One run~5 onset was recovery-censored and is
assigned only a lower-bound R6 tier.\\
\addlinespace[1pt]
Power-cycle sequence (R8) & 1 &
During run~4, boot attempts did not complete while irradiation continued.
Operation returned after a sequence that included beam termination and power
cycling. \\
\bottomrule
\end{tabular*}
\begin{tablenotes}\scriptsize
\item $N$ counts clustered operational onsets. Recovery is the least disruptive successful action when isolated; censored or multi-action episodes are reported as lower bounds or complete sequences.
\end{tablenotes}
\end{threeparttable}
\end{table}

\begin{figure*}[!p]
\centering
\begin{minipage}[t]{\textwidth}
\vspace{0pt}\centering
\includegraphics[width=\linewidth]{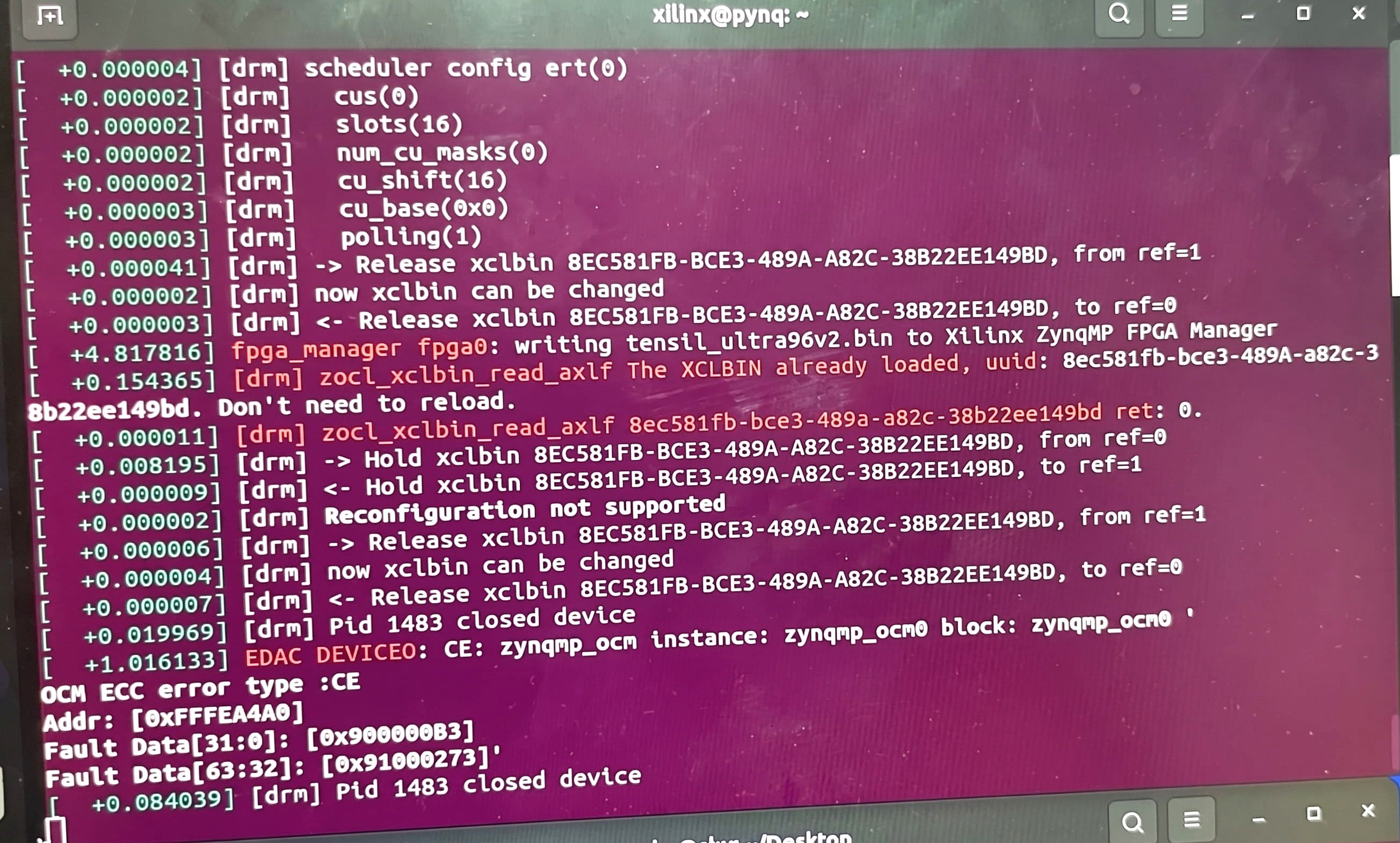}\\[-1pt]
{\footnotesize\textbf{(a)} Corrected-OCM report}
\end{minipage}

\vspace{4pt}
\begin{minipage}[b]{0.455\textwidth}
\vspace{0pt}\centering
\includegraphics[width=\linewidth]{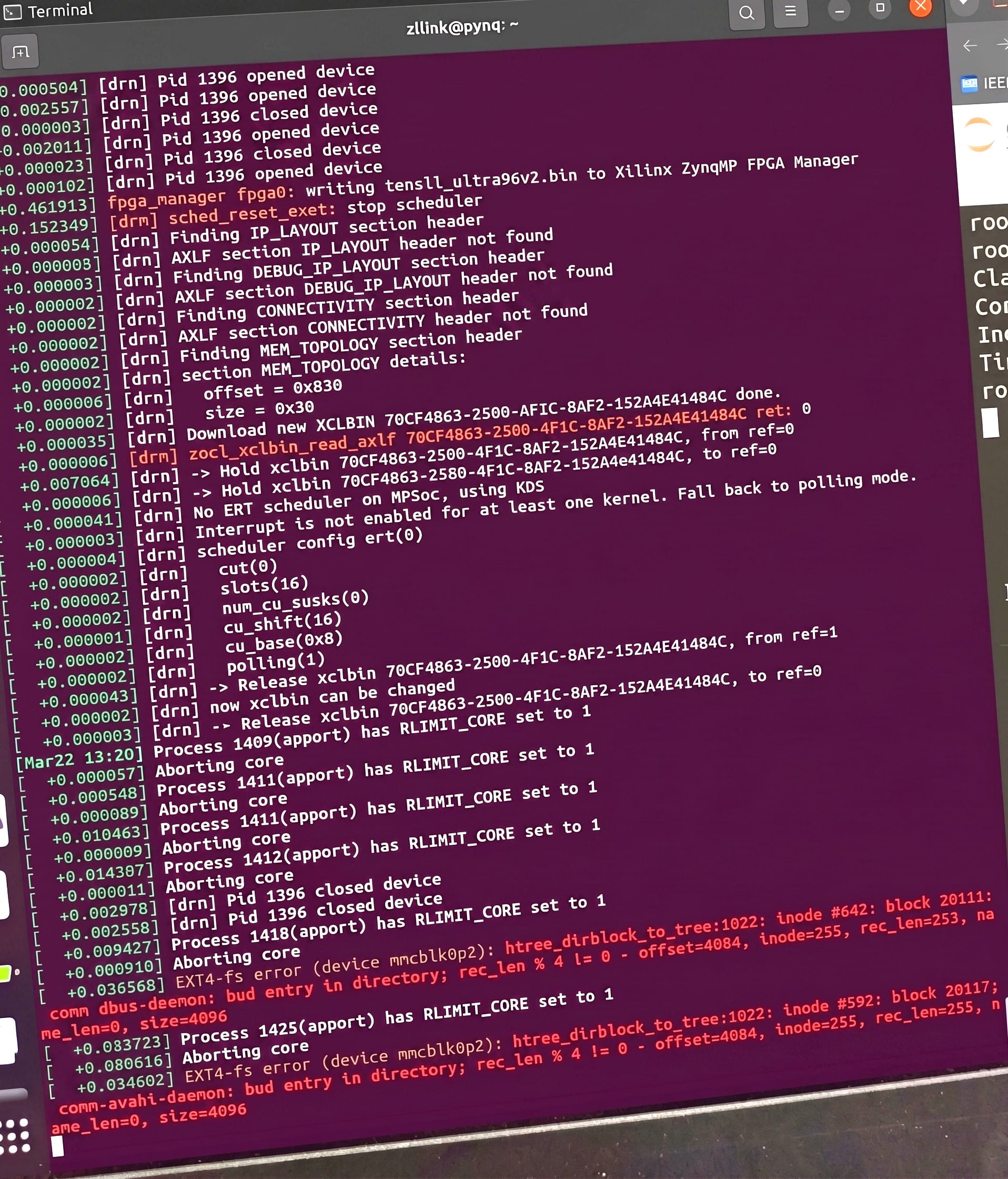}\\[-1pt]
{\footnotesize\textbf{(b)} Crash-handler failures and EXT4 errors}
\end{minipage}\hfill
\begin{minipage}[b]{0.535\textwidth}
\vspace{0pt}\centering
\includegraphics[width=\linewidth]{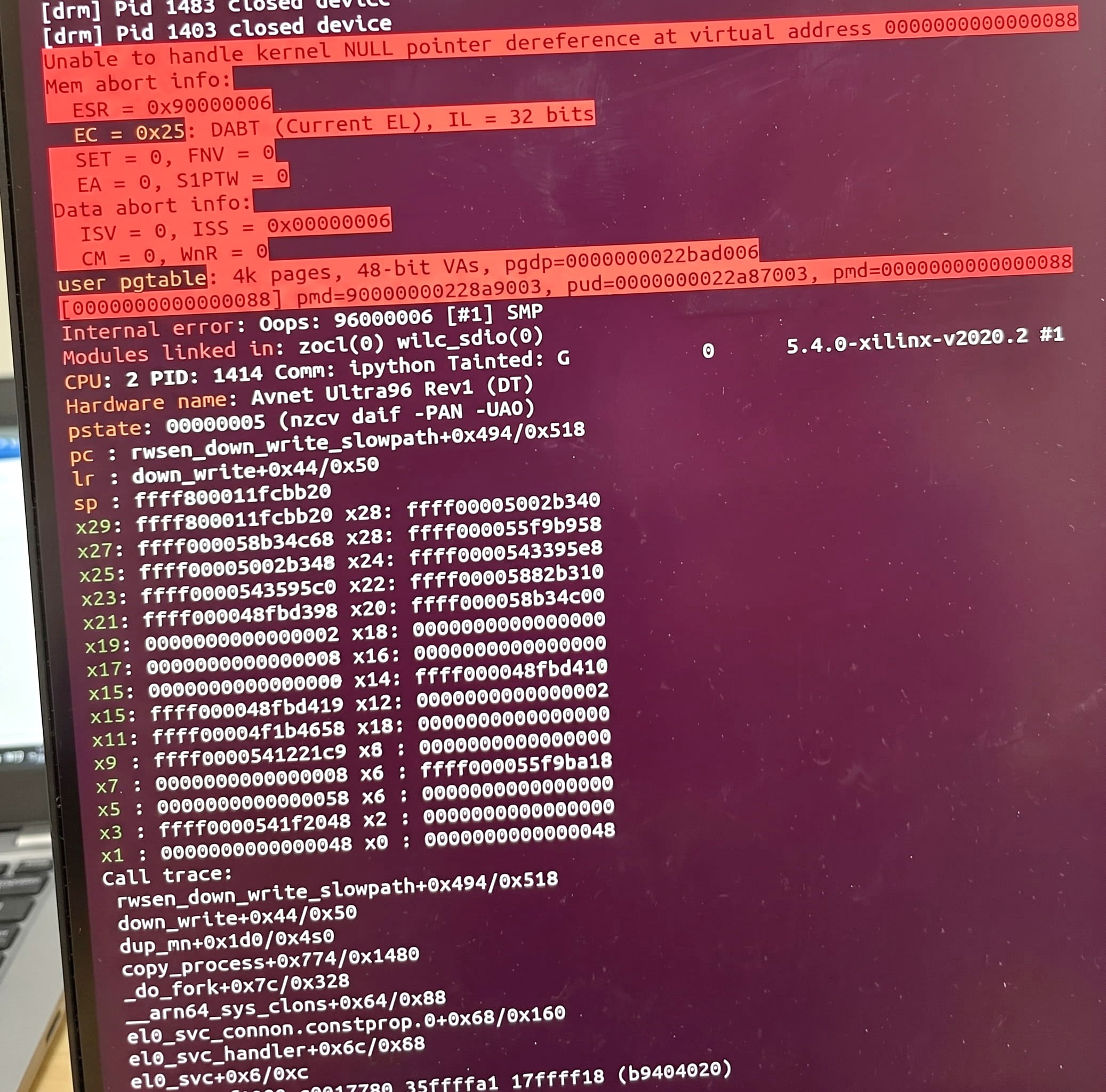}\\[-1pt]
{\footnotesize\textbf{(c)} Kernel oops and call trace}
\end{minipage}

\caption{Three archived Linux/OCM records: a corrected-OCM report, filesystem errors, and a kernel oops. Images are contrast-enhanced and checked against the originals.}
\label{fig:linuxlogs}
\end{figure*}

\begin{table}[!t]
\caption{On-Chip-Memory Correctable-Error Telemetry}
\label{tab:ocm}
\centering
\begin{threeparttable}
\footnotesize
\setlength{\tabcolsep}{3.0pt}
\begin{tabular*}{\columnwidth}{@{\extracolsep{\fill}}cccll@{}}
\toprule
Run & $E$ & Field & Address & Reporting pattern\\
& (MeV) & (cm) & & \\
\midrule
1 & 20 & 2 & \texttt{0xFFFEB8A0} & single report\\
2 & 40 & 2 & \texttt{0xFFFF6B50} & single report\\
5 & 58 & 4 & \texttt{0xFFFEA4A0} & repeated across nine blocks;\\
&    &   &                     & observed across one reboot\\
\bottomrule
\end{tabular*}
\begin{tablenotes}\scriptsize
\item All addresses lie in documented OCM \cite{ug1085}. Repeated run~5 reports at one address are grouped as one telemetry episode; that episode brackets the run~5 F4 block and one Linux-SEFI block (Fig.~\ref{fig:timeline}).
\end{tablenotes}
\end{threeparttable}
\end{table}

The seven Linux-SEFIs span three recovery levels (Tables~\ref{tab:counts} and
\ref{tab:linuxsefi}). Two F8 events stopped the prescribed workload but
cleared with a process restart while Linux remained reachable. The run~3
record contains no signal, exit code, or kernel message for the termination,
so a software-only cause cannot be excluded for that event. Of the four
reboot-tier F9 events, one is documented by a photographed console showing a
null-pointer kernel oops. Its call trace passes through the \texttt{clone}/fork
path while \texttt{ipython} was active (Fig.~\ref{fig:linuxlogs}(c)). The other
three are evidenced by loss of workload progress and notebook or secure-shell
service, subject to the USB-gadget link limitation in
Section~\ref{sec:limitations}.

Table~\ref{tab:onsets} lists the nine onsets in campaign order with the
block times shown in Fig.~\ref{fig:timeline}. The interpretation of run~4
affects the field comparison (Section~\ref{sec:partition}). Launches at 90.4
and 91.4\,min failed and were followed by a reboot. The relaunch at 97.4\,min
then failed and began the boot-failure sequence. During interval 4b, which
lasted 388\,s under beam, another launch failed within that sequence before the
beam was stopped and the board was power-cycled. We count two onsets in this
record.

$\Phi_{\mathrm{op}}$ for run~4 covers 87.4 to 91.4\,min, so the second onset
lies outside the denominator window. Crediting the minute around that relaunch
to $\Phi_{\mathrm{op}}$ would move the matched-energy $p$-value from 0.016 to
about 0.02. It would also reduce the 40-MeV wide-field cross section from
$6.9\times10^{-10}$ to $5.5\times10^{-10}$\,cm$^2$/system. Both changes fall
within the applied timing bound. We retain the defined window because the
relaunch failed within the timeline's resolution. Table~\ref{tab:sens} also
reports readings that merge the two onsets or count every failed relaunch and
credit interval 4b.

The second run~4 episode led to power cycling. Repeated boot attempts did not
complete while irradiation continued, and operation returned after a sequence
that included beam termination and a power cycle. At the wide-field onset
rates in Table~\ref{tab:xs} and each run's flux, the mean interval between
onsets is 2 to 13\,min, or 2 to 11\,min at the run~4 flux. This interval is
comparable to a Linux boot of about one minute at the 40-MeV rate, although the
episode itself contributes one of the two onsets. If the rate during boot
matched the workload rate, the onset rate alone makes repeated boot failure
under irradiation plausible. Beam termination may therefore have been
sufficient. Similar behavior has been reported on a Linux-managed ZCU102
\cite{agiakatsikas2024}. Because beam exposure and power state changed
together, the evidence supports the recorded R8 sequence but does not show that
power cycling was necessary. With only two power-rail samples per block in
runs~2--5, the record also cannot exclude a single-event latch-up or another
high-current condition during the episode.

Fig.~\ref{fig:linuxlogs} presents three archived Linux/OCM images that support
the telemetry, filesystem, and kernel summaries. The EXT4 errors in panel (b)
show similarly malformed entries in two directory blocks of two directories on
the microSD root filesystem. Both entries have the same offset and inode field
and a zero-length name. This repeated pattern is more consistent with a
storage-path or shared-state mechanism than with independent bit upsets. The
record does not show whether the corruption persisted across later reboots.
Panel (b) also records crash-handler (\texttt{apport}) failures, but the archive
does not place either record relative to a specific onset.

Corrected-OCM telemetry appeared under both fields (Table~\ref{tab:ocm}).
Runs~1 and 2 each logged a single report at a distinct address under the small
field. This is consistent with the small field reaching the SoC, although no
workload event followed over $2.36\times10^{10}\pcm$. Run~5 logged the same
address, \texttt{0xFFFEA4A0}, in nine blocks and across one reboot. We group
these reports as one episode, a deliberate exception to the onset rule. OCM
ECC corrects the data returned by a read without rewriting the stored word
\cite{ug1085}, so one stored upset is reported on every read until that word is
rewritten.

Of the nine blocks, the eight that completed retained no class record. The
ninth was either the constant-class block or a block that did not complete;
the record does not identify which. The three addresses lie in the upper OCM
region that holds the Arm Trusted Firmware image in the default boot flow. A
reboot rewrites this region, so recurrence after reboot is consistent with
either a persistent cell or repeated upsets; the record cannot distinguish
them. The 97 complete \texttt{memtester} records, comprising 1746 pattern
results with no mismatch, do not map one-to-one onto the 92 blocks. Each shows
only that a selected 10-MB region passed the 18 tests before inference. They
neither establish LPDDR4 integrity during an output event nor localize one.

\subsection{Cross Sections}
\label{sec:crosssections}

\begin{table}[!t]
\caption{Condition-Stratified Linux-SEFI Cross Sections}
\label{tab:xs}
\centering
\begin{threeparttable}
\footnotesize
\setlength{\tabcolsep}{1.5pt}
\begin{tabular*}{\columnwidth}{@{\extracolsep{\fill}}cccccc@{}}
\toprule
$E$ & Field & $N$ & $\opfl$ & $\sigma_{\mathrm{SEFI}}$ & 95\% interval or UL\\
(MeV) & (cm) & & (p/cm$^2$) & (cm$^2$/system) & (cm$^2$/system)\\
\midrule
20 & 2 & 0 & $1.19\times10^{10}$ & -- & $<2.5\times10^{-10}$\tnote{a}\\
20 & 4 & 1 & $0.78\times10^{10}$ & $1.3\times10^{-10}$ & $[3.2\times10^{-12},\,7.1\times10^{-10}]$\\
\midrule
40 & 2 & 0 & $1.17\times10^{10}$ & -- & $<2.6\times10^{-10}$\tnote{a}\\
40 & 4 & 2 & $0.29\times10^{10}$ & $6.9\times10^{-10}$\tnote{b} & $[8.4\times10^{-11},\,2.5\times10^{-9}]$\\
\midrule
58 & 4 & 4 & $0.86\times10^{10}$ & $4.7\times10^{-10}$ & $[1.3\times10^{-10},\,1.2\times10^{-9}]$\\
\bottomrule
\end{tabular*}
\begin{tablenotes}\scriptsize
\item $N$ is the clustered Linux-SEFI count; cross sections follow \eqref{eq:sigma}; -- marks an undefined point estimate for zero events. \item[a] One-sided 95\% Poisson upper limit from \eqref{eq:zero_ul}. Nonzero intervals are two-sided exact Garwood intervals. Counting, dosimetry, and bounded timing uncertainties are reported separately.
\item[b] The second onset lies outside $\Phi_{\mathrm{op}}$ (Section~\ref{sec:linuxsefi}); with the relaunch minute credited, $5.5\times10^{-10}$ $[6.7\times10^{-11},\,2.0\times10^{-9}]$.
\end{tablenotes}
\end{threeparttable}
\end{table}

\begin{table}[!t]
\caption{Output-Event Cross Sections by Field and by Nominal Energy}
\label{tab:essbit}
\centering
\begin{threeparttable}
\scriptsize
\setlength{\tabcolsep}{1.0pt}
\begin{tabular*}{\columnwidth}{@{\extracolsep{\fill}}lcccc@{}}
\toprule
Condition & $N$ & $\Phi_{\mathrm{ver}}$ & $\sigma_{\mathrm{FE}}$ & 95\% interval or UL\\
 & & ($10^{10}$\,p/cm$^2$) & (cm$^2$/system) & (cm$^2$/system)\\
\midrule
Small field (runs 1, 2) & 0 & 1.64 & -- & $<1.8\times10^{-10}$\tnote{a}\\
Wide field (runs 3--5) & 2 & 1.23 & $1.6\times10^{-10}$ & $[2.0\times10^{-11},\,5.9\times10^{-10}]$\\
\midrule
20\,MeV (both fields) & 0 & 1.93 & -- & $<1.6\times10^{-10}$\tnote{a}\\
40\,MeV (both fields) & 1 & 0.57 & $1.8\times10^{-10}$ & $[4.4\times10^{-12},\,9.8\times10^{-10}]$\\
58\,MeV (wide field) & 1 & 0.37 & $2.7\times10^{-10}$ & $[6.9\times10^{-12},\,1.5\times10^{-9}]$\\
\bottomrule
\end{tabular*}
\begin{tablenotes}\scriptsize
\item $\Phi_{\mathrm{ver}}$ follows \eqref{eq:phiver} with the verified block counts of Table~\ref{tab:runsum}; the F3 event counts as one onset (its record supports one to three). \item[a] One-sided 95\% upper limit.
\end{tablenotes}
\end{threeparttable}
\end{table}

Table~\ref{tab:xs} reports the Linux-SEFI cross sections by energy and
field. The small-field conditions yield one-sided upper limits, whereas the
wide-field point estimates have broad intervals because the counts are small.
Pooled over the wide field, seven onsets in $1.93\times10^{10}\pcm$ give
$3.6\times10^{-10}$\,cm$^2$/system
$[1.5\times10^{-10},\,7.5\times10^{-10}]$.
Table~\ref{tab:essbit} reports the output-event cross sections with the
$\Phi_{\mathrm{ver}}$ denominator of \eqref{eq:phiver}, stratified by field and
energy. Each nonzero energy-specific estimate is based on one event. The
output endpoint does not distinguish the two fields because the small-field
upper limit exceeds the wide-field estimate. Operational availability provides
a complementary view of the Linux-SEFI results. The small-field runs kept the
workload operational for 91--93\% of the beam interval; the shortfall is beam
time outside the block sequence, not downtime. Runs~3 and 5 reached 75--76\%,
and run~4 reached 21\% (Table~\ref{tab:runsum}).

The run~5 Linux-SEFI count depends on three interpretations of the record. The
alternatives merge the ambiguous Linux sequence near recovery into an adjacent
episode, exclude the recovery-censored onset, or treat the failed block after
the constant-class block as part of that episode. Each interpretation alone
reduces the 58-MeV Linux-SEFI count from four to three and its point estimate
from $4.7\times10^{-10}$ to $3.5\times10^{-10}$\,cm$^2$/system. Applied
together, they leave one or two events because the archive does not establish
whether the first two interpretations concern the same block
(Table~\ref{tab:onsets}). None changes the output-event count. The 40-MeV
wide-field value in Table~\ref{tab:xs} also halves when the run~4 events are
merged. The same numbers result if the failed launch at 90.4\,min is treated as
a continuation of the F3 event, as in the second row of Table~\ref{tab:sens}.
Treating the run~4 F3 record as three onsets would triple the 40-MeV
output-event cross section without altering the existence of class-output
corruption.

\subsection{Field Contrast}
\label{sec:partition}

\begin{table}[!t]
\caption{Sensitivity of the Matched-Energy Field Test}
\label{tab:sens}
\centering
\begin{threeparttable}
\footnotesize
\setlength{\tabcolsep}{1.5pt}
\begin{tabular*}{\columnwidth}{@{\extracolsep{\fill}}>{\raggedright\arraybackslash}p{0.52\columnwidth}ccc@{}}
\toprule
Reading of the record & Events 20/40\,MeV & $p$ & $p$ (timing)\\
\midrule
Primary: inclusive definition; recorded reboot ends a cluster & 1/2 & 0.016 & 0.043\\
Second run~4 onset merged; equivalently, the power-cycle episode excluded (JEDEC-consistent) & 1/1 & 0.079 & 0.14\\
Every failed relaunch after a recorded recovery counted; interval 4b credited to the run~4 denominator ($0.79\times10^{10}$ p/cm$^2$) & 1/3 & 0.026 & 0.050\\
Failed launch at 90.4\,min read as part of the F3 episode (alone, as the second row) and the power-cycle episode excluded & 1/0 & 0.40 & 0.46\\
Reboot-tier and power-cycle events only & 0/2 & 0.040 & 0.094\\
Reboot-tier events only & 0/1 & 0.20 & 0.31\\
Runs 1 to 4 pooled without energy stratification & 3 of 3 & 0.030 & 0.061\\
\bottomrule
\end{tabular*}
\begin{tablenotes}\scriptsize
\item One-sided exact conditional $p$-values; the timing bound lengthens every wide-field window and shortens every small-field window by its reconstruction bound. Under the primary reading the two-sided exact 95\% interval for the wide-to-small rate ratio is $[1.25,\infty)$, $[0.76,\infty)$ with the timing bound, and $[0.47,\infty)$ with the second run~4 onset merged. Crediting all run~4 exposure outside the operational window as well gives 0.066 (0.10).
\end{tablenotes}
\end{threeparttable}
\end{table}

Q2 asks whether exposing board circuitry beyond the SoC footprint changed the
Linux-SEFI rate at matched energies, not where any fault occurred. At
20\,MeV the matched exposures are $1.19\times10^{10}\pcm$ (small field) and
$0.78\times10^{10}\pcm$ (wide), with counts 0 and 1. At 40\,MeV, the exposures
are $1.17\times10^{10}$ and $0.29\times10^{10}\pcm$, with counts 0 and 2.
Under equal rates within each energy pair, \eqref{eq:conditional_p} gives
$0.78/(1.19+0.78)=0.396$ for the 20-MeV event and
$[0.29/(1.17+0.29)]^2=(0.199)^2$ for the two 40-MeV events. Because every
matched-energy onset fell in the wider field, the one-sided exact $p$-value is
their product:
\begin{equation}
 p_{\mathrm{matched}}=0.396\,(0.199)^2\approx1.6\times10^{-2}.
\label{eq:matchedfield}
\end{equation}
With only three events, this is the smallest value the test can produce.
Table~\ref{tab:sens} shows its sensitivity to alternative readings. Two of the
three events lie in the run~4 wide-field window, which lasted about four
minutes and has a $\pm150$\,s timing bound, equal to 62\% of its fluence.
Depending on how the recovery record is interpreted, this window contains one,
two, or three onsets. Across the reported readings and timing bound, $p$ ranges
from 0.016 to 0.46. The largest value excludes both run~4 interruptions from
the Linux-SEFI count, leaving one event in the test.

\section{Discussion}
\label{sec:discussion}

\subsection{Alive, on Schedule, and Wrong}
\label{sec:aliveschedule}

Run~5 shows what a reset-on-error protocol misses: the behavior of corrupted
state before the reset. The accelerator returned the same wrong class for the
remaining 39 inferences of the block, about 0.9\,s at its usual cadence. The
scheduled reload ended the observed sequence, but the next block did not
complete. The record therefore cannot establish whether the reload cleared the
underlying condition. Although the same availability indicators recorded every
availability failure in the campaign, none flagged this event. They did not
inspect output content, and the memory test and power-rail samples were taken
outside the inference phase.

The behavior is consistent with numerical or data-path corruption that leaves
control flow intact. Several fault classes could produce an input-independent
constant class at the usual cadence. These include a corrupted constant or
final-layer bias in the LPDDR4 model buffer, a stuck accumulator or SIMD path,
or an instruction stream that bypasses layers. The class index excludes a stale output buffer
(Section~\ref{sec:signatures}). The block-end model reload would refresh the
corrupted constant or bias. It would also refresh an instruction-stream
corruption if it lay in the program buffer rather than the decode logic. PL
reconfiguration would refresh the logic involved in all three candidates. The retained class sequence does not
distinguish among these candidates or exclude corruption in a transfer,
configuration path, or PS-side driver state.

These observations motivate content checks alongside availability monitors.
Each check should specify which state it validates. The top-1
oracle used here cannot detect corruption that preserves the winning class,
and the stock deployment had no content checker. Periodic known-answer inputs
could bound an episode to the probe interval if their state and expected
outputs were protected. A class-histogram check would have flagged the absent
class at its first occurrence. Checksums of static model buffers before each
reload, together with a bounded wait and DMA status check, could test the
candidates above at low cost. None of these measures was evaluated here.

\subsection{Persistent State and Recovery Scope}

The campaign recorded output corruption, a repeatedly reported corrected-error
address, and malformed directory entries. Only the output events were silent.
The longer output event lasted for the rest of its block without an alert from
any availability indicator. It may have continued as the subsequent hang, a
sequence observed after both output events. If so, the condition resided in
state that the block-end reload did not refresh (Section~\ref{sec:procedure}).
This would exclude the three candidates in Section~\ref{sec:aliveschedule} and
point instead to state beyond the reach of a process restart or PL reload. The
corrected-error address was reported in nine blocks spanning a reboot,
consistent with a persistent cell or repeated upsets. However, ECC corrected
every read and telemetry logged every report, so the corrupted value was never
delivered to software. The malformed entries appeared in two root-filesystem
directory blocks at one time. They replicate one pattern but do not show
persistence, and their origin is unknown.

Each recovery action refreshes different state. A process restart recreates
userspace state but leaves Linux, drivers, PS state, external memory, and PL
configuration unchanged. In this protocol, however, every recovery was
followed by a new block that re-downloaded the PL and reloaded the model. The
protocol therefore refreshed more state than the recovery action alone. On
this platform, a Linux reboot performs a system reset that
clears PL configuration and reruns the boot firmware \cite{ug1085}. PYNQ loads
no bitstream at boot, so the next block start re-downloaded the PL. A reboot
still cannot remove a persistent hardware condition; the run~5 OCM address
returned after one. PL reconfiguration with a model reload restores the
implemented logic, initialized PL memories, and model buffers, but not Linux,
OCM, other LPDDR4 contents, or the page cache. Power cycling resets the
broadest set of state at the cost of the longest interruption. SEM scrubbing,
had it been used, would have covered PL configuration memory but not block-RAM
contents, flip-flops, PS registers, OCM, LPDDR4, or software state
\cite{pg187}.

Recovery should therefore escalate according to the state that may remain
corrupted. A process supervisor can restart a terminated notebook. A failed
known-answer check can trigger a model reload or PL reconfiguration, followed
by further escalation only if the refreshed system fails again. This guidance
follows from the refresh analysis, not from a comparison of recovery methods;
no recovery strategies were tested against each other.

\subsection{What the Field Contrast Supports}
\label{sec:fieldsupport}
The field comparison supports an association, not component-level
attribution. All nine primary onsets and all three matched-energy Linux-SEFIs
occurred under the wider field, which received 45.0\% of the operational
exposure. Under the small field, corrected-OCM reports were consistent with
exposure of the SoC (Section~\ref{sec:linuxsefi}), yet no Linux-SEFI occurred in 44
completed blocks and no output event occurred in the 31 verified blocks. This
pattern is consistent with a contribution from circuitry outside the SoC
footprint and supports treating the stack, rather than the chip, as the
observed system. It does not identify the responsible circuitry. Every
wide-field run followed both small-field runs, so field is confounded with run
order and accumulated dose. The 58-MeV run also has no small-field counterpart.

Table~\ref{tab:compare} summarizes the monitoring scope and reported outcomes
of selected Linux-managed Zynq irradiation campaigns.

\begin{table}[!t]
\caption{Selected Linux-Managed Zynq Irradiation Campaigns}
\label{tab:compare}
\centering
\begin{threeparttable}
\scriptsize
\setlength{\tabcolsep}{2.3pt}
\begin{tabular*}{\columnwidth}{@{\extracolsep{\fill}}>{\raggedright\arraybackslash}p{0.29\columnwidth}
>{\raggedright\arraybackslash}p{0.64\columnwidth}@{}}
\toprule
Study and platform & Monitoring scope and reported outcomes\\
\midrule
This work; protons; Ultra96-V2, ZU3EG; PYNQ Linux + Tensil &
Workload progress/recovery, Linux records, reachability, sampled rails, and
retained top-1 classes (62 of 92 blocks). Seven Linux-SEFI and two
class-changing output-event onsets under the clustering rule; class-preserving
changes were outside the oracle.\\
\addlinespace[1pt]
Stirk \emph{et al.} \cite{stirk2023}; neutrons; Ultra96; Linux + Dhrystone &
Reported 68 unexpected SoC failures, each followed by a power cycle: 13
process-failure/hang events, 17 kernel failures, and 38
application-processing-unit resets. Its uniform power-cycle recovery differs
from the staged recovery used here.\\
\addlinespace[1pt]
Sabogal \emph{et al.} \cite{sabogal2019recon}; neutrons; UltraZed-EG, ZU3EG;
PetaLinux + ReCoN CNN &
Heartbeat liveness and golden checksums per output. For the simplex
accelerator (scrubbed static region, DDR ECC), 25 erroneous and one hung
execution in 75,527 executions at
$3.49\times10^{11}$ n/cm$^2$; consecutive erroneous outputs were counted as one
error. Class sequences and Linux failure classes were not reported.\\
\addlinespace[1pt]
Agiakatsikas \emph{et al.} \cite{agiakatsikas2024}; neutrons; ZCU102,
XCZU9EG; Linux + Vitis DPU &
Among 5985 ResNet-50 runs, reported 2964 correct, 89 crashes, 46 critical silent data corruptions (SDCs)
associated with misclassification, and 2886 tolerable SDCs without a final-class change. Its numerical-output checker detects effects that a top-1-only
checker cannot observe.\\
\bottomrule
\end{tabular*}
\begin{tablenotes}\scriptsize
\item Counts are not susceptibility rankings. Particle spectra, devices, workloads, monitor coverage, clustering, recovery rules, and denominators differ.
\end{tablenotes}
\end{threeparttable}
\end{table}

\subsection{Limitations and Threats to Validity}
\label{sec:limitations}

The campaign evaluated a specific hardware--software configuration: a Tensil
accelerator implemented on an Ultra96-V2 and running a compiled ResNet-20
workload on a fixed set of ten CIFAR-10 images. The resulting cross sections
apply only to this system and protocol; they do not quantify device-to-device
variation, susceptibility across neural-network models, CIFAR-10 accuracy
under irradiation, or flight qualification. The two beam configurations also
did not isolate individual components. The edge of the nominal 2-cm field may
have reached the LPDDR4 package, whereas the nominal 4-cm field exposed
additional board resources. Run order was not randomized, and every wide-field
run followed the small-field runs. Five unmonitored same-day exposures added
dose outside the reported denominators, and 58\,MeV was tested only under the
wide field. Consequently, the association between wide-field
exposure and the observed onsets cannot be attributed specifically to LPDDR4.
The field comparison remains exploratory, and the energy-stratified cross
sections in Tables~\ref{tab:xs} and \ref{tab:essbit} do not establish an
energy dependence.

The experiment did not retain logits, intermediate tensors, model-memory
checksums, DMA transactions, physical LPDDR4 addresses, or PL configuration
readback. It therefore cannot reconstruct the internal propagation path of
either output event or localize any onset. Because Linux evidence reached the
laptop through the USB-gadget link, loss of notebook or secure-shell service
cannot be distinguished from loss of the link itself without a local console.
A link loss would have appeared as a workload interruption and could affect
both the onset count and the assigned tier. The userspace memory
test covered only 10\,MB of LPDDR4 and did not retain physical-page mappings,
limiting both memory coverage and fault localization.

\section{Conclusion}
\label{sec:conclusion}

Proton irradiation exposed two failure modes in this Linux-managed inference
stack. Seven Linux-SEFIs stopped the workload. Two output-corruption events
instead returned incorrect classes while execution continued. The longer event
showed a stuck-class pattern: 39 consecutive CIFAR-10 inputs were assigned the
same class, which was absent from the ten-image pool. Outputs continued at the
usual 23--24\,ms interval, and the availability monitors reported no anomaly.
The stuck-class sequence was still present when the scheduled block-end
bitstream reconfiguration began. The next launch failed, so the record cannot
determine whether reconfiguration cleared the underlying condition. The other
output event was also followed by a failed launch. The kernel also reported
corrected OCM ECC
errors at the same address in nine blocks spanning one reboot. ECC corrected
every read, and the record does not link these reports to the output corruption.

All nine event onsets in the primary analysis occurred under the wider beam
field. This pattern is consistent with a contribution from circuitry outside
the SoC footprint. However, field size was confounded with run order and
accumulated dose. The results therefore do not localize the failures or
identify LPDDR4 as their source.

These results show that liveness and nominal timing do not establish inference
correctness. Linux-managed accelerators need end-to-end, content-aware checks
and staged recovery that can refresh the layers where corrupted state may
persist.
Radiation campaigns should count events by onset, match each cross-section
denominator to the monitor that can observe its endpoint, and state that
monitor's blind spots. Recovery actions should be reported without using them
to infer the fault's physical origin. Future tests should retain complete score
vectors, add a local console, randomize the beam-field order, and include long
runs without scheduled bitstream reloads. Such runs would allow fault
persistence to be measured beyond one block. This study provides a
system-level baseline for future software hardening of COTS FPGA-SoCs used for
neural-network inference in modern space systems.

\section*{Acknowledgment}
The authors thank the operations staff of the Department of Radiation Research
and Proton Radiotherapy at IFJ PAN for beam time and dosimetry support. OpenAI
ChatGPT (GPT-5.6 Pro, September 2026) and Anthropic Claude (Fable 5.1,
September 2026) were used for language editing, structural revision, and
consistency checking of the abstract and Sections~I--VI. They did not generate
experimental observations. The authors checked all numerical values,
equations, citations, interpretations, figures, and final wording against the
campaign records and accept full responsibility for the manuscript.

\end{document}